\documentclass[a4paper, amsfonts, amssymb, amsmath, reprint, showkeys, nofootinbib, superscriptaddress]{revtex4-2}
\usepackage[english]{babel}
\usepackage[utf8]{inputenc}
\usepackage[colorinlistoftodos, color=green!40, prependcaption]{todonotes}
\usepackage{bm}
\usepackage{color}
\usepackage[unicode=true,pdfusetitle,bookmarks=true,
bookmarksnumbered=false,bookmarksopen=false,breaklinks=false,
pdfborder={0 0 0},backref=false,colorlinks=true,citecolor=red]
{hyperref}
\usepackage{cleveref}
\usepackage{amsthm}
\usepackage{mathtools}
\usepackage{physics}
\usepackage{graphicx}
\usepackage[left=23mm,right=13mm,top=35mm,columnsep=15pt]{geometry} 
\usepackage{adjustbox}
\usepackage{placeins}
\usepackage[T1]{fontenc}
\usepackage{lipsum}
\usepackage{csquotes}
\usepackage{graphicx}
\usepackage{subcaption}
\usepackage[normalem]{ulem}

\begin{document}

\begin{abstract}
Boundary conditions dictate how fluids, including liquid crystals, flow when pumped through a channel. Can boundary conditions also be used to control internally driven \emph{active} fluids that generate flows spontaneously? By using numerical simulations and stability analysis we explore how surface anchoring of active agents at the boundaries and substrate drag can be used to rectify coherent flow of an active polar fluid in a 2D channel. Upon increasing activity, a succession of dynamical states is obtained, from laminar flow to vortex arrays to eventual turbulence, that are controlled by the interplay between the hydrodynamic screening length and the extrapolation length quantifying the anchoring strength of the orientational order parameter. We highlight the key role of symmetry in both flow and order and show that coherent laminar flow with net throughput is only possible for weak anchoring and intermediate activity. Our work demonstrates the possibility of controlling the nature and properties of active flows in a channel simply by patterning the confining boundaries.
\end{abstract}

\title{Boundaries control active channel flows}
\date{\today} 
\author{Paarth Gulati}
\affiliation{Physics Department, University of California, Santa Barbara, California}
\author{Suraj Shankar}
\affiliation{Physics Department, Harvard University, Cambridge, Massachusetts}
\author{M.~Cristina Marchetti}
\affiliation{Physics Department, University of California, Santa Barbara, California}
\maketitle

\maketitle
\section{Introduction}
\label{sec:intro}

Active fluids are composed of active entities, such as bacteria \cite{wensink2012meso,zhou2014living}, biofilaments driven by motor proteins \cite{sanchez2012spontaneous,schaller2010polar} or self-propelled colloids \cite{palacci2013living,bricard2013emergence}, that consume energy to generate their own motion. Such active particles exert dipolar forces on their surroundings, driving self-sustained  active flows. The elongated nature of the active units endows the fluid with liquid crystalline degrees of freedom, allowing for the onset of orientational order, with either polar or nematic symmetry \cite{marchetti2013hydrodynamics}.

The orientationally ordered state of bulk active fluids is generically unstable at all activities due to the feedback between deformations and flow~\cite{simha2002hydrodynamic}, resulting in spatiotemporal chaotic dynamics at zero Reynolds number that has been referred to as bacterial or active turbulence~\cite{alert2021active}. Several strategies have been proposed to stabilize laminar flows in active fluids such as the inclusion of substrate friction~\cite{duclos2014perfect,doostmohammadi2016stabilization,duclos2018spontaneous,thijssen2021submersed} or spatial confinement~\cite{voituriez2005spontaneous,wioland2013confinement,lushi2014fluid,wu2017transition,chen2018dynamics,opathalage2019self,you2021confinement}, but a systematic treatment unifying these various results has remained elusive.

The dynamics of confined two-dimensional (2D) active fluids also depends on the symmetry of local order, though some features are common.
Polar active fluids, such as dense suspensions of swimming bacteria, transition from laminar to undulating and periodic travelling flows upon increasing the channel width, eventually giving place to turbulent dynamics~\cite{wioland2016directed,tjhung2011nonequilibrium,giomi2008complex,yang2016role}. In active nematics, both numerical studies \cite{chandragiri2019active,shendruk2017dancing,samui2021flow,wagner2022exact} and experiments with microtubule-kinesin suspensions \cite{hardouin2019reconfigurable} with strong anchoring to the channel walls have revealed a transition from laminar to oscillatory flows to a lattice of counter-rotating flow vortices with associated order of disclinations in the nematic texture. Similar flow states and transitions are also reported in other geometries, such as in circular confinement \cite{wioland2013confinement,opathalage2019self,norton2018insensitivity}. Recent work has also begun exploring the influence of varying and conflicting anchoring boundary conditions in active nematics in channels \cite{yang2016role,rorai2021active}. In general the interplay of the geometry of confinement and boundary conditions yields a rich variety of flow states, but coherent flow with a finite throughput in the channel is only achieved by finely tuning activity and other system parameters.
Hence, quantifying the conditions that yield  specific flow patterns, and especially identifying states of finite throughput is important for controlling bacterial flow through channels and for microfluidic applications of active flows \cite{poujade2007collective, conrad2018confined, beebe2002physics, clark2015modes,needleman2017active}.

\begin{figure}[h]
    \centering{
    \includegraphics[width=0.95\linewidth]{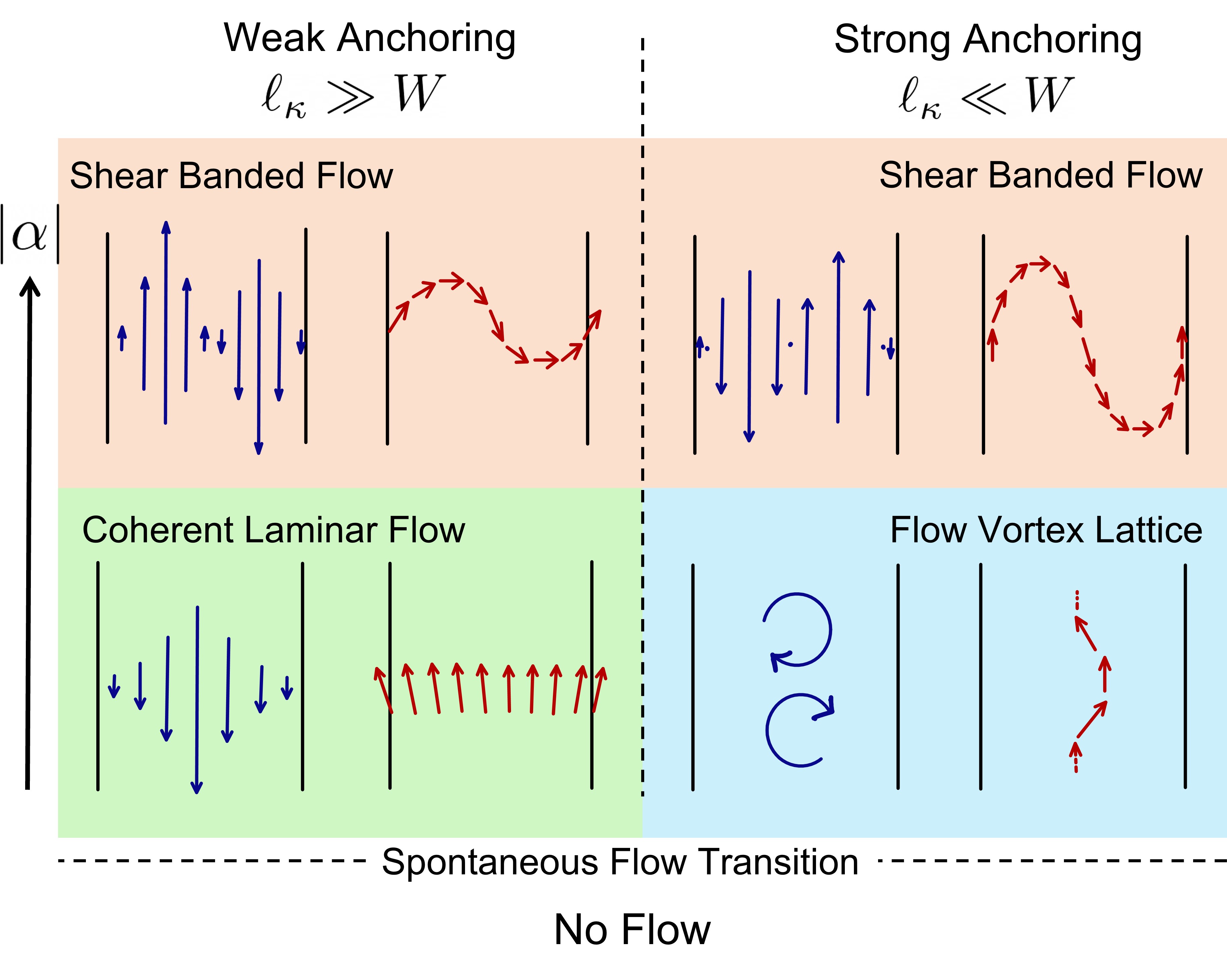}}
    \caption{Schematic showing the effect of wall anchoring of the polarization on the steady flow states of a polar active fluid in a channel of width $W$. 
    The blue arrows depict the flow field, $\mathbf{v}$, and the red arrows represent the polarization, $\mathbf{p}$. The extended arrow on the left indicates the direction of increasing activity $\abs{\alpha}$.}
    \label{fig:FlowSketches}
\end{figure}

Motivated by the sensitivity of liquid crystals to surface effects and anchoring on the boundary \cite{degennes}, in this paper, we suggest a simple strategy to control channel flows of active fluids by tuning boundary conditions.
We consider an incompressible polar active fluid confined to a two-dimensional (2D) channel with friction, and examine the role of surface anchoring in selecting the spontaneously flowing states of the fluid. The model may be appropriate, for instance, to describe the spontaneous flow of a bacterial suspension in a channel. We show that the selection of the flow patterns is controlled by the interplay of two length scales: (i) the hydrodynamic screening length 
$\ell_\eta=\sqrt{\eta/\Gamma}$ that quantifies the scale beyond which dissipation by substrate friction ($\Gamma$) dominates over dissipation from internal shear viscosity ($\eta$), and (ii) the extrapolation length, $\ell_\kappa=K/E_a$ governing the length scale over which elastic torques controlled by the stiffness $K$ of the polar fluid balance the wall anchoring energy $E_a$ of the orientational degrees of freedom. The hydrodynamic length screens flows and controls the penetration of the boundary conditions on the flow field $\mathbf{v}$ into the bulk of the channel. The extrapolation length controls the relative strength between nematic elasticity and surface anchoring. Strong wall anchoring corresponds to a short $\ell_\kappa$, while weak anchoring corresponds to large $\ell_\kappa$
~\cite{degennes}.

We find that coherent active flows with finite throughput are only possible when the orientational order parameter is weakly anchored to the channel walls, corresponding  to large  $\ell_\kappa$ compared to the width $W$ of the channel.  
In the opposite limit of strong anchoring, the spontaneous flow transition leads instead to a single file of flow vortices evenly spaced along the length of the channel that we refer to as flow vortex lattice. These vortices appear in counter-rotating pairs and their number is determined by the aspect ratio of the channel and the activity.
The succession of flow states obtained when activity is increased above the spontaneous flow instability on the way to turbulence are summarized schematically in Fig.~ \ref{fig:FlowSketches} for both weak and strong anchoring. By considering variable boundary conditions, we unify previous results in a comprehensive phase diagram that crucially combines the well-known active length scale $\ell_\alpha=\sqrt{K/|\alpha_0|}$ ($\alpha_0$ measuring the strength of the active stress) \cite{giomi2015geometry,hemingway2016correlation} controlling patterns in the bulk of the fluid along with boundary related length scales in both flow ($\ell_\eta$) and order ($\ell_\kappa$).

In the rest of the paper, we first introduce the hydrodynamic model and the boundary conditions used in the channel geometry. In section~\ref{sec:numerics} we report results from numerical solutions of the continuum equations and describe the various spontaneous flow states observed with increasing activity. We define and evaluate the mean normalized throughput through the channel to distinguish between coherent and non-coherent flows. In section~\ref{sec:stability} we present  a linear stability analysis of the hydrodynamic model for a rectangular periodic box that qualitatively accounts for the transitions between the various flow states. The results are summarized in a comprehensive phase diagram in terms of activity, hydrodynamic screening and anchoring strength. 
In Section~\ref{sec:polar} we discuss the effect of a polar propulsive force in the momentum equation which breaks flow symmetry and destroys the non-flowing ordered state. Finally, we conclude with a discussion of potential experimental realisations and possible extensions of our work. 

\section{Hydrodynamic Model}
\label{sec:model}

We consider a two-dimensional active polar fluid  on a frictional substrate, as appropriate, for instance, to describe a thin film of a bacterial suspension~\cite{maitra2020swimmer}. At high bacterial concentration, we assume both the suspension density and the bacterial concentration  to be constant and describe the dynamics in terms of two fields, the bacterial polarization $\mathbf{p}$ that characterizes the local direction of bacterial motility and the flow velocity $\mathbf{v}$ of the fluid. 

The dynamics of the polarization is governed by 
\begin{equation}
D_t\mathbf{p} =  \lambda \mathbf{S}\cdot \mathbf{p}+\frac{1}{\gamma}\mathbf{h} \;,
\label{eq:p}
\end{equation}
where
$D_t\mathbf{p} = \partial_t\mathbf{p} + \mathbf{v}\cdot\bm\nabla\mathbf{p} + \boldsymbol\Omega\cdot\mathbf{p}$ is the material derivative that embodies advection and rotation of polarization by flow, with  $\boldsymbol\Omega =\left(\bm\nabla \mathbf{v} - {\bm\nabla \mathbf{v}}^T\right)/2$ the  vorticity tensor.  The first term on the right-hand side of Eq.~\eqref{eq:p} describes flow alignment, with $\mathbf{S}= \left(\bm\nabla \mathbf{v} + {\bm\nabla \mathbf{v}}^T \right)/2$ the strain rate tensor and $\lambda$ a microscopic parameter that depends on the shape of the active entities ($\lambda>1$ for elongated swimmers). The second term is the molecular field that drives relaxation with a rate set by the rotational viscosity $\gamma$. It is determined by a Landau free energy as $\mathbf{h} = -{\delta F}/{\delta \mathbf{p}}$,
with
\begin{equation}
F= \frac12 \int_\mathbf{r}\left\{K(\partial_ip_j)^2  - a\left(\frac{c}{c_0}-1\right) \mathbf{p}^2 + \frac{b}{2}\mathbf{p}^4\right\}\;,
\label{eq:F}
\end{equation}
where $a,b>0$ and the bacterial concentration $c$ controls the transition to polar order. Here, we have assumed that a single elastic constant $K$ controls the stiffness to both bend and splay distortions. For simplicity we neglect in Eq.~\eqref{eq:p} flow alignment terms proportional to $\mathbf{v}$ which arise through a lubrication approximation~\cite{maitra2020swimmer}. We have verified that these terms do not qualitatively change our results.

At low Reynolds number the flow is governed by force balance through the Stokes equation,

\begin{equation}
    \Gamma\mathbf{v}= - \boldsymbol{\nabla}\Pi+
    \eta \nabla^2 \mathbf{v} + \boldsymbol{\nabla}\cdot\big( \boldsymbol{\sigma}^\text{a} +  \boldsymbol{\sigma}^\text{lc}\big)\;,
    \label{eq:stokes}
\end{equation}
where the pressure $\Pi$ is determined by the condition of  incompressibility,
$\bm\nabla\cdot\mathbf{v} =0$.   
Here,  $\Gamma$ is the friction with the substrate. Dissipation is controlled by the interplay of friction and  viscosity $\eta$,  with $\ell_\eta=\sqrt{\eta/\Gamma}$ the viscous screening length that controls the penetration of the no-slip boundary conditions into the channel. The passive liquid-crystalline stress, $\boldsymbol\sigma^\text{lc}$,
describes the elastic stresses due to  distortions of the polarization field and is given by
\begin{equation}
\boldsymbol\sigma^\text{lc} = -\frac{\lambda +1}{2}\mathbf{p}\mathbf{h} - \frac{\lambda -1}{2}\mathbf{h}\mathbf{p} + K \boldsymbol\nabla p_i \boldsymbol\nabla p_i\;.
\label{eq:sigmalc}
\end{equation}
Finally, the dipolar forces exerted by the swimmers on the fluid yield an active stress $\boldsymbol\sigma^\text{a}$ \cite{marchetti2013hydrodynamics}, 
\begin{equation}
\bm\sigma^\text{a}= \alpha_0\left(\mathbf{p}\mathbf{p}- \frac{\abs{\mathbf{p}}^2}{2}\mathbf{I}\right)\;,
\label{eq:sigmaa}
\end{equation}
where the activity $\alpha_0$ provides a measure of the strength of active forces, depending, for instance, on bacterial concentration and swimming speed. Its sign depends on whether such forces are extensile ($\alpha_0<0$ as for pushers) or contractile ($\alpha_0>0$ as for pullers). Here we focus on extensile active forces which are relevant to most bacteria. Note that, other sources of activity such as self-advection are neglected here for simplicity, and their effects are briefly discussed later in Sec.~\ref{sec:polar}

We assume that the fluid is confined to a channel of width $W$ and length $L$, in the geometry shown in Fig.~\ref{fig:CorrelationFunc}, with periodic boundary conditions along the $y$ direction.  Below we focus on the dynamics of the homogeneous ordered state with   $c > c_0$ and  $\mathbf{p}=\sqrt{a/b}~\mathbf{\hat{y}}$. We  normalize the polarization so that $|\mathbf{p}|=1$ in the aligned state.

The hydrodynamic equations are solved with the boundary conditions
\begin{eqnarray}
 &&\mathbf{v} \big|_{x=0, W} =0\;, \label{eq:noslip}\\
&&\left[ E_a (\mathbf{p} - \hat{\mathbf{y}})+ K (\hat{\mathbf{n}}\cdot{\bm \nabla}\mathbf{p})\right] _{x=0, W} =0\;,
\label{eq:polarAnchoring}
\end{eqnarray}
where $\hat{\mathbf{n}}$ is a unit normal pointing outward from the walls. 
The boundary condition on the polarization expresses the balance between a torque $E_ap_x$ that penalizes misalignment with the boundary, with $E_a$ an anchoring energy per unit length, and the nematic torque $K\partial_xp_x$ that penalizes deviations from the aligned state.  The ratio {$\ell_\kappa=K/E_a$} defines the extrapolation length \cite{degennes}, with the following limiting cases
\begin{equation}
\begin{aligned}
\ell_\kappa\rightarrow 0:~~&\left[\mathbf{p}\right]_{x=0,W}=\mathbf{\hat{y}}~~~{\rm strong~anchoring}
\notag\\
\ell_\kappa\rightarrow \infty:~~&\left[\partial_x\mathbf{p}\right]_{x=0,W}=\mathbf{0}~~~{\rm weak~anchoring}
\end{aligned} 
\end{equation}
The hydrodynamic equations for our model have nematic symmetry, as they are invariant for $\mathbf{p}\xrightarrow{} -\mathbf{p}.$ The boundary conditions, however, break the symmetry in polarization by aligning $\mathbf{p}$ with the channel walls in $\hat{\mathbf{y}}$ direction. As we will discuss later in Sec \ref{sec:polar}, this still allows for symmetry in the flow direction, which can by broken by introducing an active self propulsion term in Eq.~\ref{eq:stokes}.

\section{Numerical Simulations}
\label{sec:numerics}

\begin{figure}[h]
    \centering
    \includegraphics[width=0.9\linewidth]{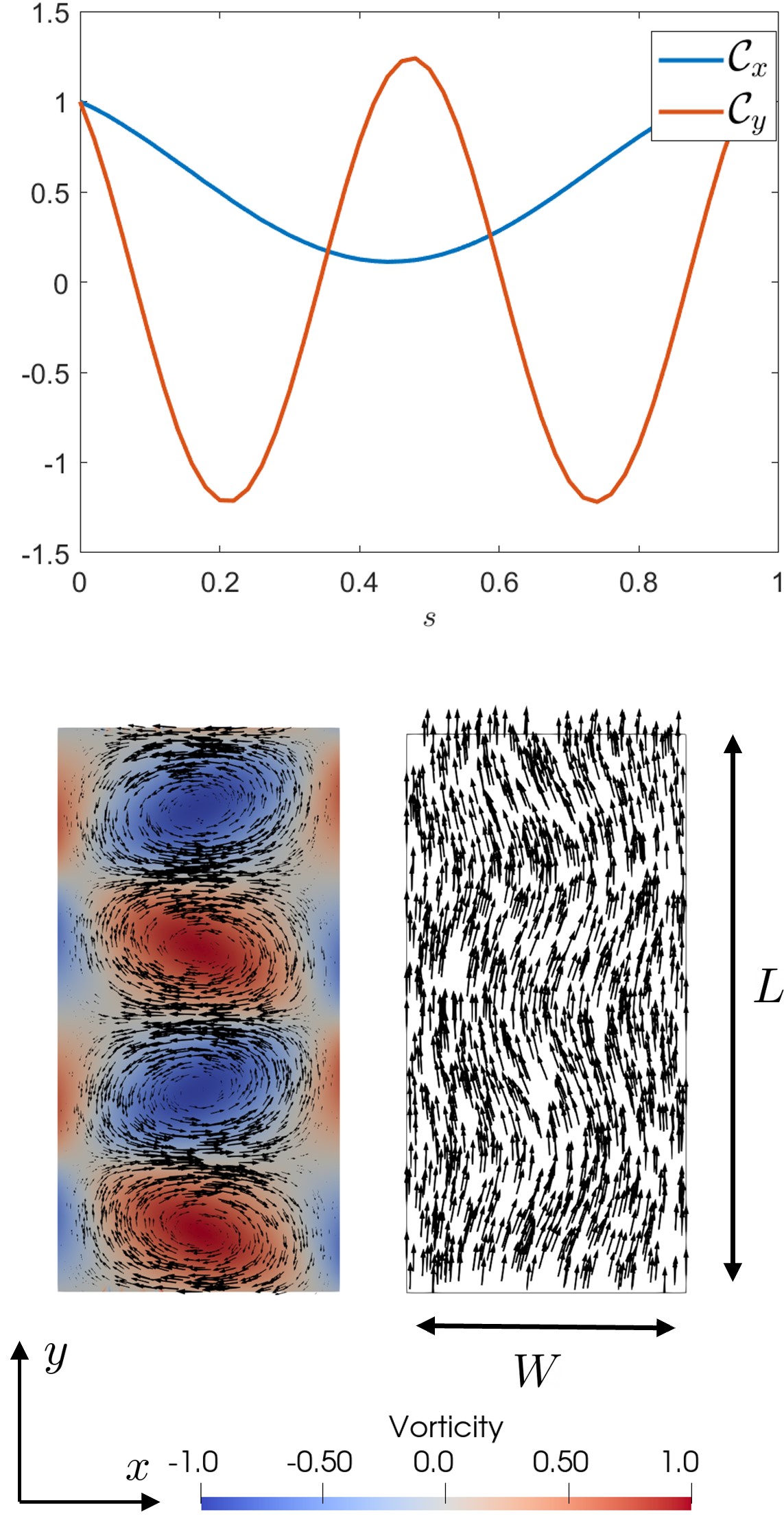}
    \caption{The velocity correlation function across (along) the channel, $C_\perp$ ($C_\parallel$) for the vortex lattice state ($\ell_\eta =0.30, \alpha =-2.00$) with strong anchoring ($\ell_\kappa =0.01$). The correlation is plotted against $s$, the nondimensionalized distance across (along) the channel in terms of $x/W (y/L)$, respectively. The flow profile $\mathbf{v}$ (bottom left) and polarization $\mathbf{p}$ (bottom right) in the vortex lattice state are plotted at the bottom. }
    \label{fig:CorrelationFunc}
\end{figure}

\begin{figure*}[ht]
    \centering
    \begin{subfigure}[b]{0.48\linewidth}
        \centering
        \includegraphics[width=\textwidth]{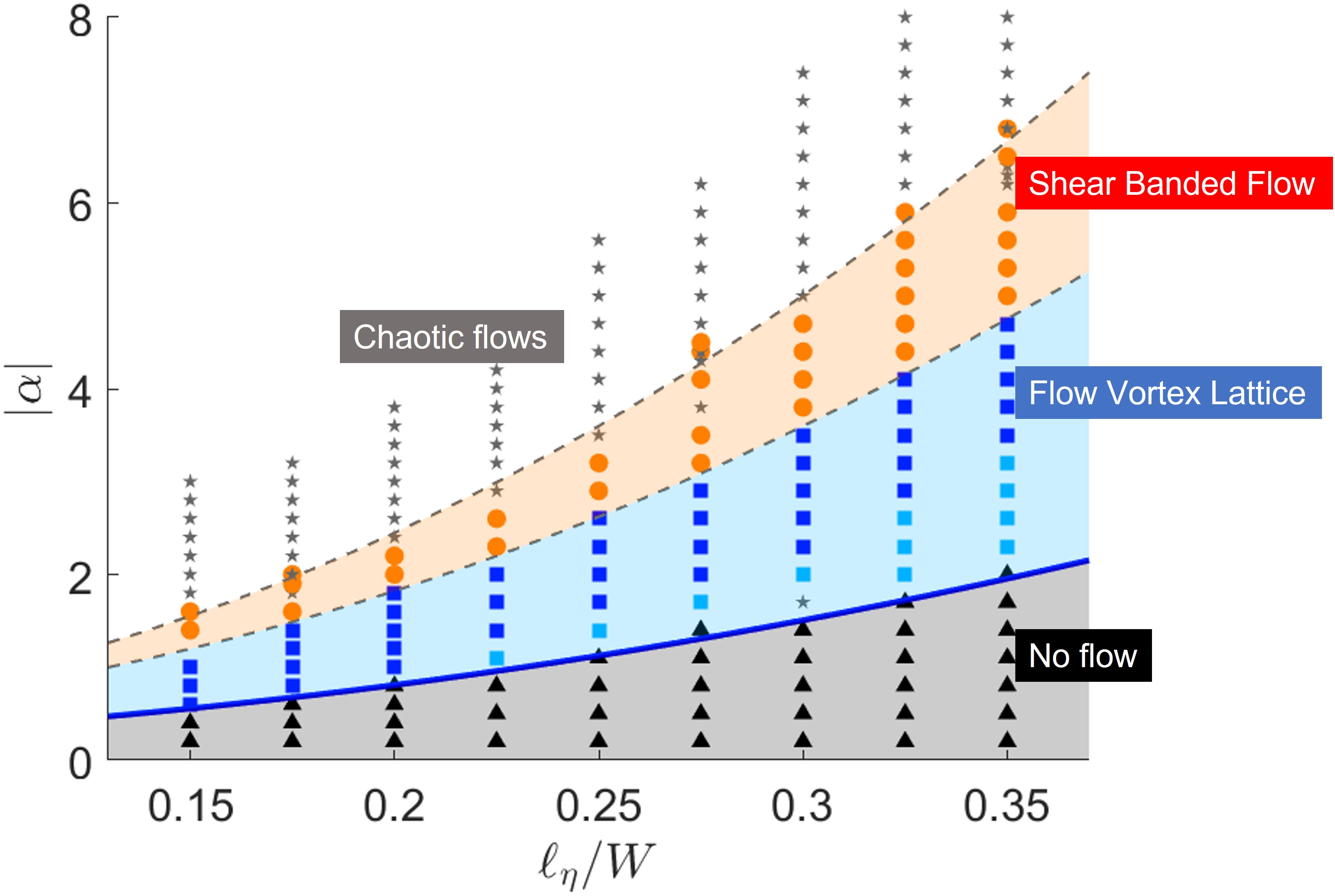}
        \caption{Strong anchoring, $\ell_\kappa/W = 0.01$}
        \label{fig:PhaseStrongAnchor}
     \end{subfigure}
     \hfill
     \begin{subfigure}[b]{0.50\linewidth}
        \centering
        \includegraphics[width=\textwidth]{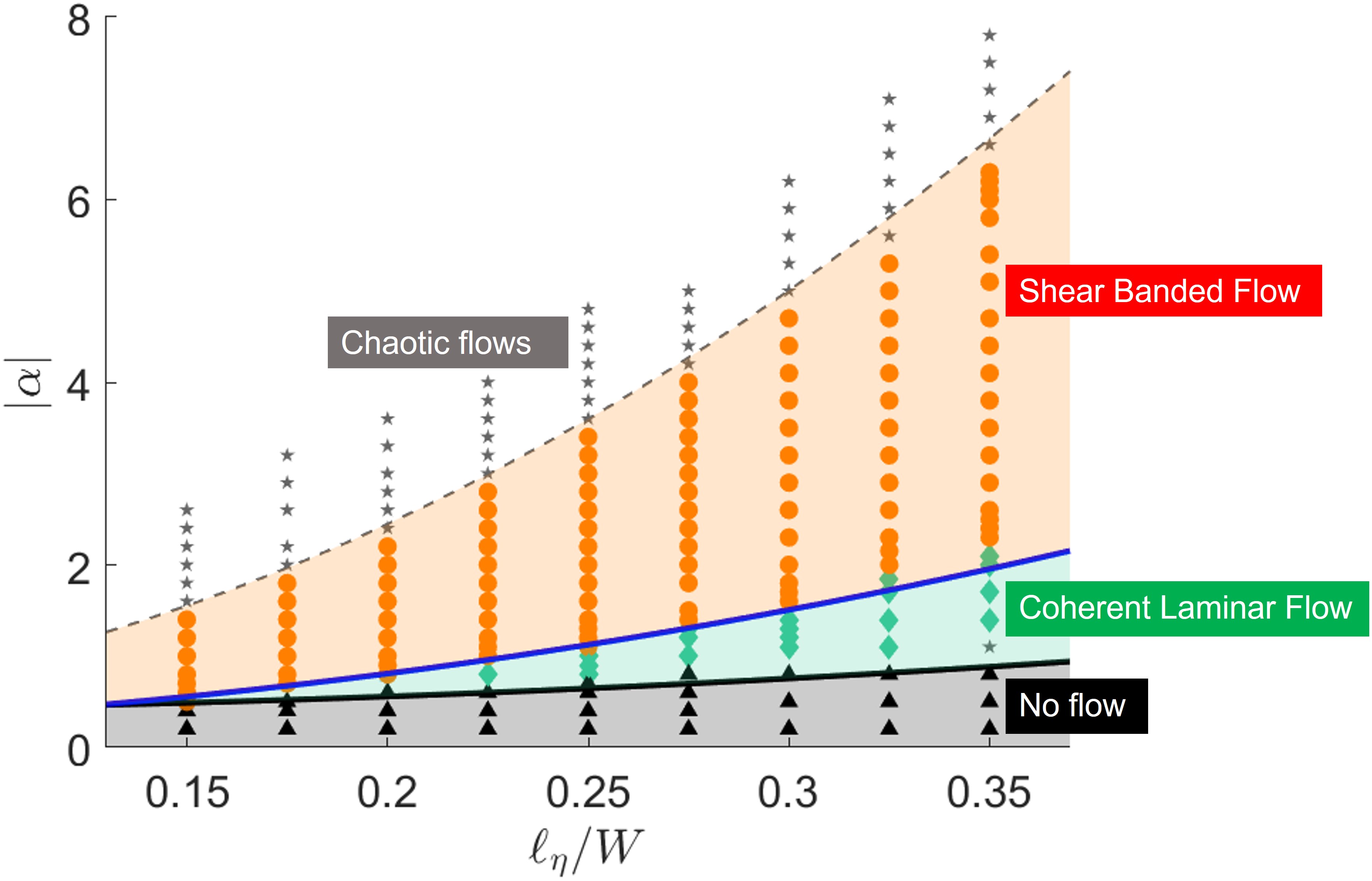}
        \caption{Weak anchoring, $\ell_\kappa/W=100$}
        \label{fig:PhaseWeakAnchor}
    \end{subfigure}
    \caption{These phase diagrams show the various steady flow states in the channel as we change the screening length and the activity at a fixed anchoring strength. The dashed lines are fits to the observed phase boundaries whereas the solid lines correspond to the various curves calculated using linear stability analysis in Sec. \ref{sec:stability}, with no fitting parameters. The solid blue line in both phase diagrams corresponds to the \emph{mixed} bend-splay instability $\alpha^m$ and the solid black line in \ref{fig:PhaseWeakAnchor} corresponds to the \emph{splay} instability with weak anchoring, $\alpha^{s,w}$.}
    \label{fig:}
\end{figure*}
\begin{figure*}[ht]
    \centering
    \begin{subfigure}[b]{0.45\linewidth}
         \centering
         \includegraphics[width=\textwidth]{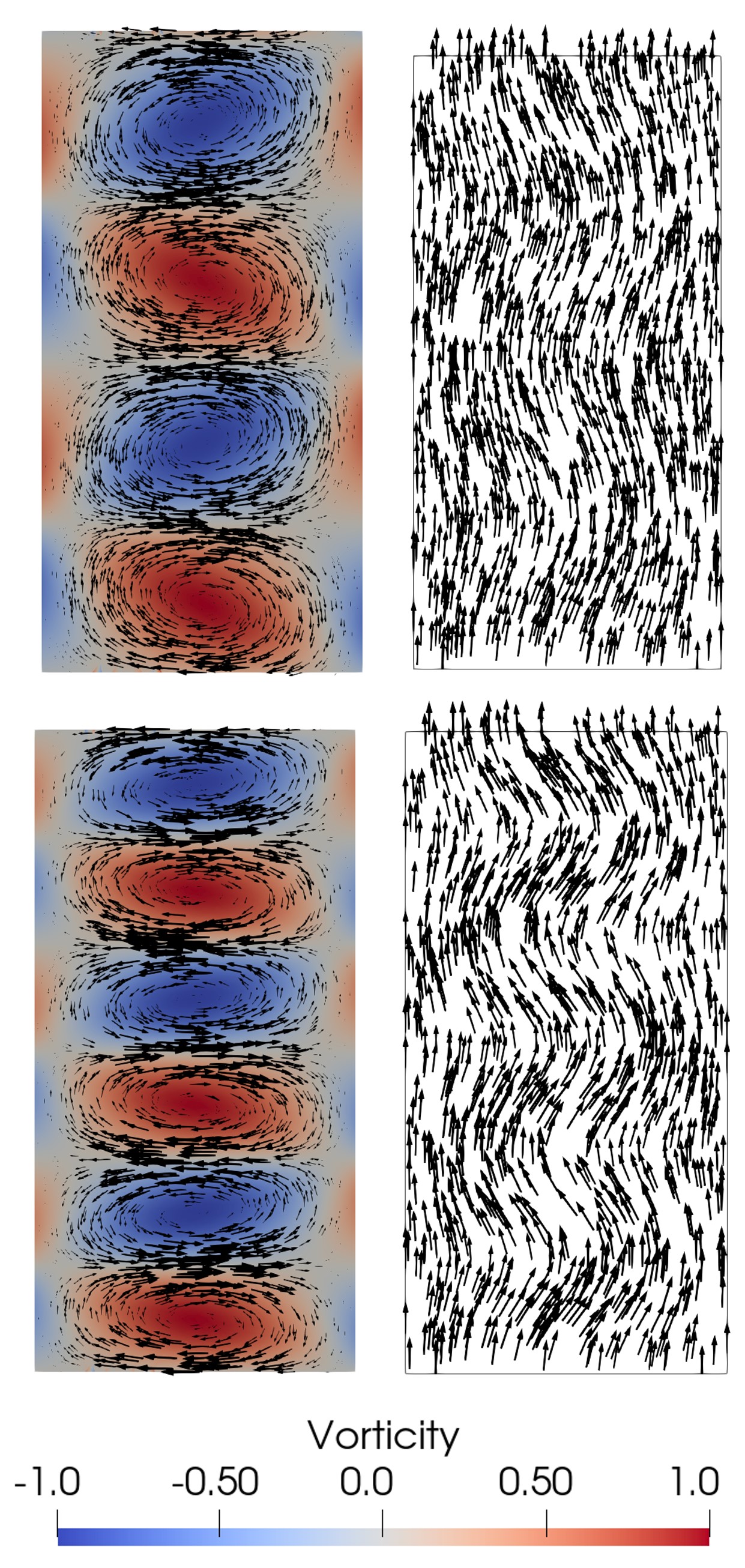}
         \caption{Steady states with strong anchoring, $\ell_\kappa/W = 0.01$, and low activity (top, $\alpha =-2.00$) and high activity (bottom, $\alpha =-3.00$)}
         \label{fig:Profiles_Strong}
     \end{subfigure}
     \hfill
     \begin{subfigure}[b]{0.45\linewidth}
         \centering
         \includegraphics[width=\textwidth]{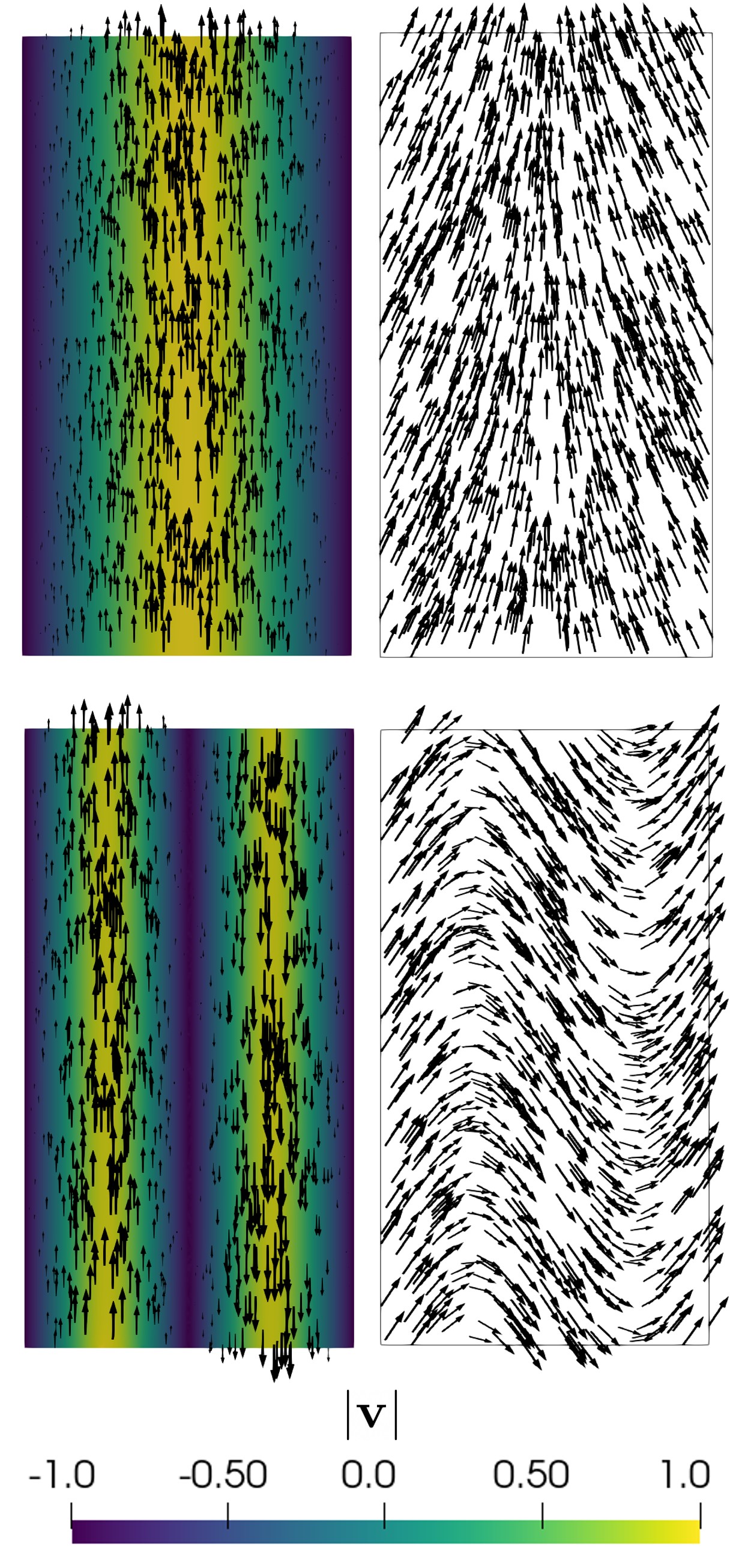}
         \caption{Steady states with weak anchoring, $\ell_\kappa/W = 100$, and low activity (top, $\alpha =-1.75$) and high activity (bottom, $\alpha =-4.00$)}
         \label{fig:Profiles_Weak}
     \end{subfigure}
     \caption{Some of the steady states observed in the phase diagrams ($\ell_\eta =0.35$) are shown in terms of the flow $\mathbf{v}$ (left in each frame) and the polarization $\mathbf{p}$ (right). The vorticity and the flow magnitude are colorized to show relative magnitudes and have been scaled by the corresponding maximum/minimum. The arrows represent the vectors for $\mathbf{v}$ and $\mathbf{p}$ scaled by their magnitudes.}
      \label{fig:}
\end{figure*}

\begin{figure*}[ht]
    \centering
    \includegraphics[width=0.9\linewidth]{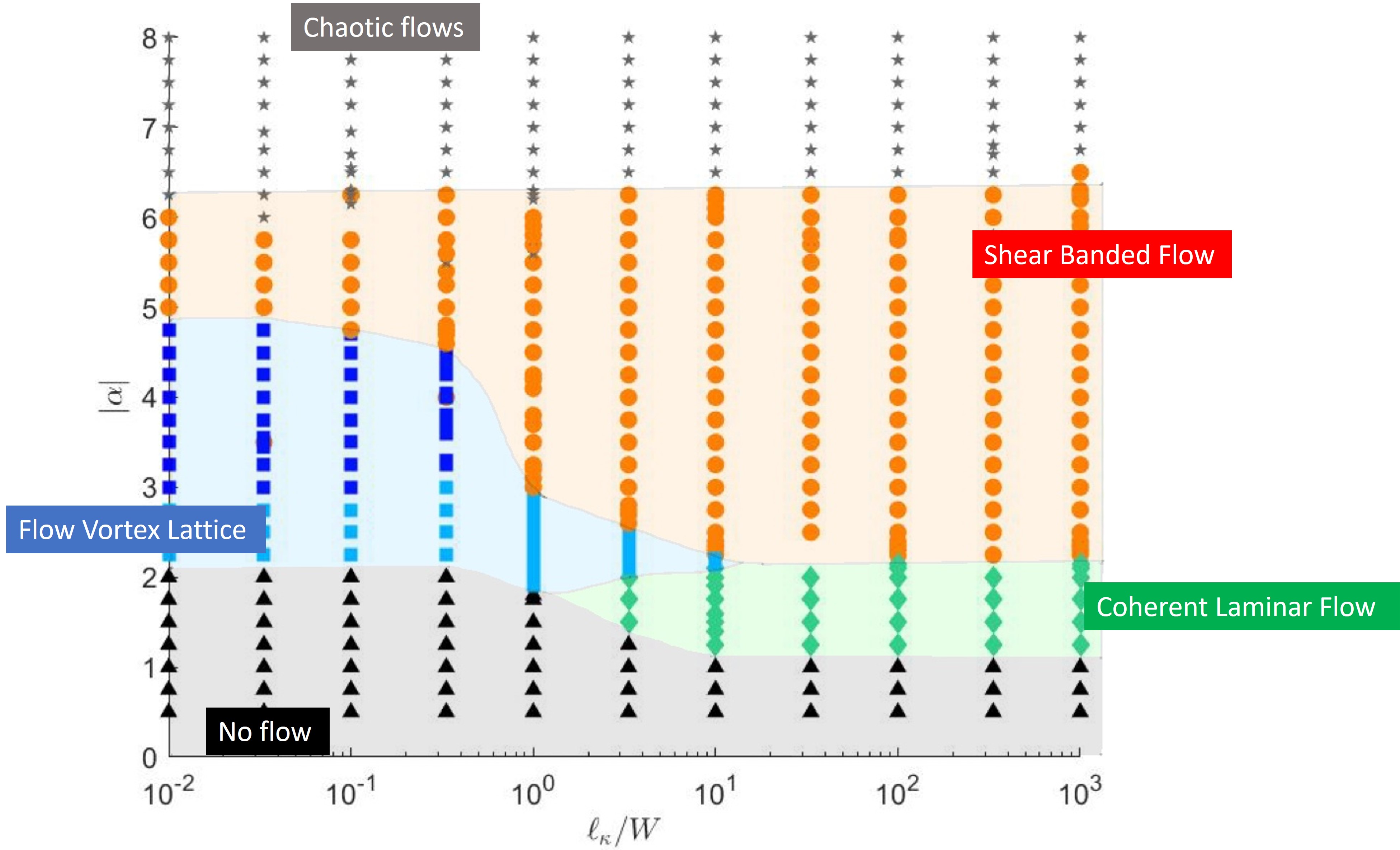}
    \caption{Phase diagram as we change the extrapolation length $\ell_\kappa$ at a fixed screening length $\ell_\eta =0.35$}
    \label{fig:PhaseDiagramAnchoringStrength}
\end{figure*}

The hydrodynamic equations are solved numerically using the finite element platform FEniCs \cite{alnaes2015fenics,logg2012automated}. We use the width of the channel, $W$, as unit of length, the nematic relaxation time, $\tau_n = \gamma/a$, as unit of time
and  the condensation energy $a$ in the polar free energy as a unit of stress. 
With this choice, the dimensionless activity $\alpha=|\alpha_0|/a=(\ell_\alpha/\xi)^2$ is simply the square of the ratio between the active length $\ell_\alpha=\sqrt{K/|\alpha_0|}$ and the nematic correlation length $\xi=\sqrt{K/a}$. 
Unless specified otherwise, the results are shown for channel dimensions with $L =2W$, flow alignment parameter  $\lambda = 2$, and  nematic correlation length  $\xi=0.1W$.
To carry out  the finite element simulations, the channel is triangulated into a rectangular mesh with grid size $\mathrm{d}x \sim 0.01 W$. The time step used in most simulations is $\mathrm{d}t = \tau_n/10$ and we typically run simulations for a total time $T=1000\tau_n$ or longer.

In bulk, an extensile active fluid ordered along $y$,  is  destabilized at any value of activity by the unbounded growth of bend fluctuations  $\delta p_x(y,t)$  of the order parameter~\cite{simha2002hydrodynamic,voituriez2005spontaneous}. Both substrate friction ($\Gamma$) and a finite system size ($L$ along the $y$ direction) generate a finite for the onset of spontaneous flow given by~\cite{duclos2018spontaneous}
\begin{equation}
{\alpha}_{c}^b= \frac{2K\Gamma}{\gamma (\lambda+1)}\left[1+ \left(\frac{ 2\pi \ell_\eta}{L}\right)^2 \right]\;.
\label{eq:bendcritical}
\end{equation}
For vanishing friction ($\Gamma\to 0$), the screening length diverges ($\ell_\eta\to\infty$) and we recover the viscous limit of the finite size activity threshold $\alpha^L_c=(2K\eta/\gamma(\lambda+1))(2\pi/L)^2$ \cite{voituriez2005spontaneous}.

As shown below, wall anchoring geometrically frustrates this bulk instability mode and alters the mechanisms and nature of the instability which is now controlled by an interplay of three length scales: the channel width $W$, the flow screening length $\ell_\eta$ and the extrapolation length $\ell_\kappa$.

Upon increasing activity, the quiescent state in a channel is destabilized, driving the system through a succession of flowing states. 
To classify such dynamical states, we examine the velocity correlation functions along and parallel to the channel, defined as
\begin{equation}
\begin{aligned}
C_{\perp}(x) = \langle \mathbf{v}(\mathbf{r})\cdot \mathbf{v}(\mathbf{r} + x\mathbf{\hat{x}}) \rangle_{\mathbf{r}}\;,   
\notag\\
C_{\parallel}(y) = \langle \mathbf{v}(\mathbf{r})\cdot \mathbf{v}(\mathbf{r} + y\mathbf{\hat{y}}) \rangle_{\mathbf{r}}\;,
\end{aligned} 
\end{equation}
where $\langle\cdot\rangle_{\mathbf{r}}$ denotes a spatial average over the entire channel domain. The number of oscillations in the correlation functions is used to classify the nature of the flow; see Fig.~\ref{fig:CorrelationFunc} for an example of a flow state (the vortex lattice) with the associated correlation functions plotted.

While $C_{\parallel,\perp}$ capture vorticity and shear in flow patterns, states with finite throughput are quantified by evaluating the normalized throughput 
\begin{equation}
\phi = \left|\left\langle\frac{v_y}{\abs{\mathbf{v}}}\right\rangle_{\mathbf{r}}\right|\;,
\label{eq:thru}
\end{equation}
where $\langle\abs{\mathbf{v}}\rangle_{\mathbf{r}}$ is the mean velocity. 

\paragraph{Strong anchoring: $\ell_\kappa \ll W$.}

The flow states obtained for strong anchoring (here $\ell_\kappa/W = 0.01$) are summarized in the phase diagram of Figs.~\ref{fig:PhaseStrongAnchor} obtained by varying the activity $\alpha$ and the screening length $\ell_\eta$. Upon increasing activity at fixed ${\ell}_\eta$, we first observe a transition from a quiescent ordered state (grey region of Fig.~\ref{fig:PhaseStrongAnchor}) to a flowing state (blue region) with $\langle\abs{\mathbf{v}}\rangle\ne 0$. Strong anchoring suppresses pure bend fluctuations. The instability is then controlled by a growing mode that necessarily has both splay and bend components, as discussed in Section~\ref{sec:stability}. This is evident from the polarization profiles displayed in Fig.~\ref{fig:Profiles_Strong}. The resulting flow is a lattice of counter-rotating flow vortices that span the channel width, with zero net throughput. We refer to this state as a ``vortex lattice''.

The number $n$ of counter-rotating vortex pairs in the vortex lattice  is controlled by the aspect ratio $L/W$ of the channel, together with the topological constraint that the net vorticity must be zero. This can be understood by comparing the energy cost of bend and splay deformations transverse and parallel to the long direction of the channel, with the number of pairs of vortices $n\sim(L/W)(A_s/A_v)$, where $A_s$ and $A_v$ are the amplitude of the polarization angle in the vortex-lattice and the shear banded flow states (Appendix \ref{sec:appendixElasticEnergy}). The minimum number of pairs of counter-rotating vortices is equal to the integer part of the aspect ratio and as we increase activity, additional vortices are added in pairs (light and dark blue points in Fig.~\ref{fig:PhaseStrongAnchor}).

Upon further increasing the activity, we note a transition (blue to orange in Fig.~\ref{fig:PhaseStrongAnchor}) from the ordered vortex lattice to a shear banded flow. This state is characterized by a bend about the short channel direction, transverse to the original orientation of the polarization field. Here, the flow and polarization are invariant along the channel. Interestingly, the flow transitions to a `more ordered state' on increasing activity. 

\paragraph{Weak anchoring, $\ell_\kappa \gg W$}
For weak anchoring, and sufficiently large values of $\ell_\eta$, the steady state corresponds to coherent laminar flow with finite throughput (green region in Fig~\ref{fig:PhaseWeakAnchor}). But as we increase activity, an ordered vortex lattice similar to that in the strong anchoring limit forms transiently but then leads to a shear banded flow as the long time steady state (Fig.~\ref{fig:Profiles_Weak}).

Notably, the laminar flowing state has non-zero \emph{splay} but zero bend
and can be seen only above a characteristic screening length $ \ell_\eta^*$ (dependant on the extrapolation length $\ell_\kappa$). From Fig.~\ref{fig:PhaseWeakAnchor}, this value $\ell_\eta^* \approx 0.2W$ for $\ell_\kappa/W= 100$. This absence of coherent laminar flow at short screening length is further discussed in Sec.~\ref{sec:stability}.

Fig.~\ref{fig:PhaseDiagramAnchoringStrength} shows a phase diagram as a function of activity and extrapolation length,  at a fixed screening $\ell_\eta= 0.35 > \ell_\eta^*$.  
As $\ell_\kappa$ decreases the coherent flow region for the splay state vanishes and the ordered flow vortices become stable for larger activities. We find that the transition to the flow vortex lattice (transient in the weak anchoring limit) is largely unaffected by the anchoring strength. Also, we observe that the activity threshold for the onset of unsteady chaotic flows is not affected by the extrapolation length, generalizing previous results \cite{opathalage2019self,norton2018insensitivity}.

\section{Linear Stability Analysis}
\label{sec:stability}
The steady state channel flows summarized in the previous section can be understood using a linear stability analysis of an initial uniformly aligned state with no flow.

First, let us consider the stability of an unconfined active fluid on a frictional substrate. For small activity, the uniform quiescent state with finite polarization $\mathbf{p}=\mathbf{\hat{y}}$ and zero flow is stable. In the absence of friction, this state is generically unstable for any activity~\cite{simha2002hydrodynamic,voituriez2005spontaneous}. The presence of friction yields a finite activity threshold for the onset of spontaneous flow~\cite{duclos2018spontaneous}. A linear stability analysis of the homogeneous state shows that fluctuations in the Fourier amplitude of wavevector $\mathbf{q}$  of the transverse component of polarization $\delta p_x$ evolve as $\delta p_x(\mathbf{q},t)\sim e^{\nu(\mathbf{q})t}$, with the growth rate
\begin{equation}
  \begin{aligned}
    \nu(\mathbf{q}) = -\frac{K}{\gamma} {q}^2+  \frac{\alpha_0({q}_x^2-{q}_y^2)}{2\Gamma(1+ \ell_\eta^2{q}^2)}\left[\lambda(\hat{q}_x^2-\hat{q}_y^2)-1\right]+\mathcal{O}(q^4)\;,
  \end{aligned}
  \label{eq:omegaLSA}
\end{equation}
where, $q= \abs{\mathbf{q}}$, $\hat{q}_{x,y}=q_{x,y}/q$  and $\ell_\eta=\sqrt{\eta/\Gamma}$ is the viscous screening length.\\

\paragraph{Bulk.}
For extensile systems ($\alpha_0<0$) of elongated active units ($\lambda>1$), the decay rate given in Eq.~\eqref{eq:omegaLSA} can become positive, signalling the instability of the homogeneous state. 
It is known that in a bulk system, defined as one with periodic boundary conditions in all directions, the 
most unstable modes are bend deformations of the polarization field, corresponding to spatial variations along the direction of order, i.e., $q_x=0$ and finite $q_y$.
The hydrodynamic instability sets in at the longest wavelength, which in a periodic box of size $L$ is $2\pi/L$, yielding an activity threshold $\alpha_c^b$ (Eq.~\ref{eq:bendcritical}).
Note, here the threshold is defined by its absolute value. The bend modes of wavelength $2\pi/L$ become unstable as $\alpha_0 < -\alpha^b_c$ or $\abs{\alpha_0} > \alpha^b_c$.

Viscous dissipation enters through the screening length $\ell_\eta$ and shifts the instability to higher values of activity whenever $\ell_\eta\sim L$. In other words, viscous dissipation stabilizes the uniform quiescent state. When $\ell_\eta\ll L$ one recovers the infinite system frictional threshold,
${{\alpha}}_{c}^\infty = 2K\Gamma/[\gamma (\lambda+1)]$.  The bulk result given in Eq.~\eqref{eq:bendcritical} fits very well with finite element simulations in a periodic box (not shown). 

Upon further increasing activity, splay modes, corresponding to $q_y =0$, also become unstable. This  splay instability occurs above a threshold 
\begin{equation}
{\alpha}_{c}^s= \frac{2K\Gamma}{\gamma (\lambda-1)}\Big[1+ \Big(\frac{ 2\pi\ell_\eta}{W}\Big)^2 \Big]\;.
\end{equation}
Note the dependence on the width $W$ of the channel rather than the length $L$.
In a periodic square box ($L=W$), for elongated flow-aligning swimmers ($\lambda >1$), $\alpha_c^s>\alpha_c^b$, i.e., the bend instability always precedes the splay instability in an extensile system \cite{marchetti2013hydrodynamics}.
The situation can, however, be reversed for large values of $\ell_\eta$ and suitable aspect ratios ($W/L$).

\begin{figure}[h]
    \centering
    \begin{subfigure}[b]{0.23\textwidth}
         \centering
         \includegraphics[width=\textwidth]{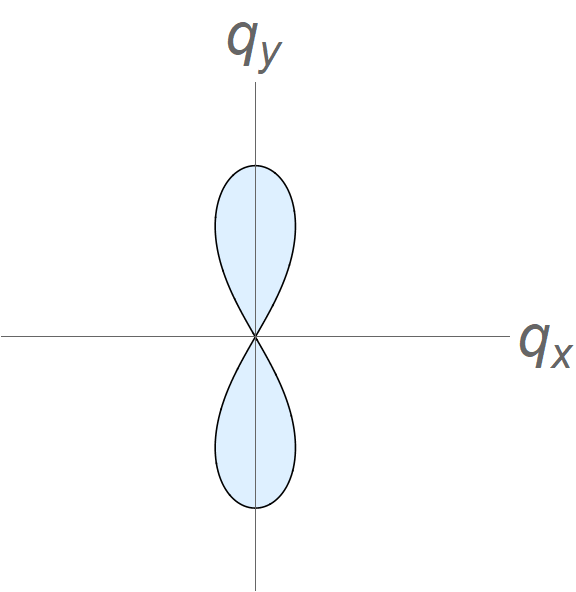}
         \caption{$\alpha_c^b < \abs{\alpha_0} <  \alpha_c^s $}
         \label{}
     \end{subfigure}
     \hfill
     \begin{subfigure}[b]{0.23\textwidth}
         \centering
         \includegraphics[width=\textwidth]{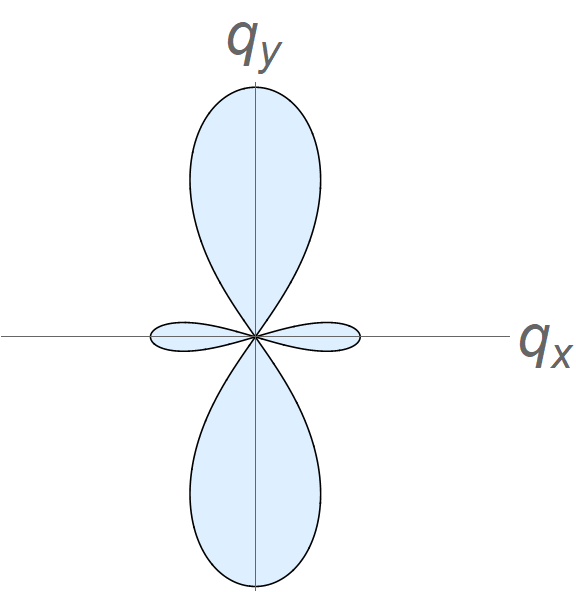}
         \caption{$ \abs{\alpha_0} > \alpha_c^s$}
     \end{subfigure}
     \caption{The shaded region in the $q_x$-$q_y$ plane corresponds to unstable modes ($\nu > 0$) in an extensile active fluid. Beyond the critical activity $\alpha_c^b$ only the bend modes ($q_x =0$) become unstable. Further increasing the activity beyond $\alpha_c^s$, pure splay modes ($q_y =0)$ also become unstable.}
      \label{fig:LSALobes}
\end{figure}

The angular dependence of the instability threshold is displayed in the polar plots of  Fig.~\ref{fig:LSALobes}, where the shaded region corresponds to unstable modes ($\nu(\mathbf{q}) >0$) in the $q_x-q_y$ plane. It is evident that the fastest growing modes are always along the $q_x=0$ direction and become unstable for $|\alpha_0|> \alpha_c^b$. Pure splay modes (corresponding to $q_y=0$) are stable for $|\alpha_0|<\alpha_c^s$ (Fig.~\ref{fig:LSALobes}(a)) and only become unstable for 
$|\alpha_0|>\alpha_c^s$, as evident  by the emergence of the two additional lobes elongated along the $q_x$ axis (Fig.~\ref{fig:LSALobes}(b)).   

\paragraph{Channel.}
\noindent
In a channel geometry, boundary conditions can differentially frustrate bend and splay distortions \cite{green2017geometry,rorai2021active}, allowing for distinct modes of spontaneous flow transitions to emerge.
As in Ref.~\cite{duclos2018spontaneous}, we first consider a quasi-1D model that assumes only spatial variations along the direction $x$ of the channel width.  This is consistent with the observation that in the channel bend fluctuations ($q_y\not=0$) are suppressed either by 
strong anchoring requiring $p_x=0$ at the boundaries~\cite{voituriez2005spontaneous,duclos2018spontaneous} and/or by the condition of no-slip. The instability to spontaneous flow is then controlled by splay fluctuations ($q_x\not=0$) and the threshold activity generally depends on the anchoring length, $\ell_\kappa$.  For strong anchoring  ($\ell_\kappa\rightarrow 0$) the no-slip requirement on the velocity further excludes the possibility of a mode with wavelength $2W$. Hence the longest allowed wavelength corresponds to $q_x=2\pi/W$ and the splay (s) instability threshold for the case of strong anchoring (a) can be estimated as 
\begin{equation}
{{\alpha}}_c^{s,a}  = \frac{2K\Gamma}{\gamma (\lambda-1)}\Big[1+  \left(\frac{2\pi\ell_\eta}{W}\right)^2 \Big]\;, \qquad \ell_\kappa = 0\;.
\label{eq:alphac-strong}
\end{equation}
For weak anchoring ($\ell_\kappa\rightarrow\infty$) conversely $\partial_xp_x=0$ at the boundaries, allowing for a cosine wave of wavelength $2W$ instead that also satisfies no-slip. Hence the splay (s) instability threshold for the case of weak anchoring (w) is estimated as
\begin{equation}
{{\alpha}}_c^{s,w} = \frac{2K\Gamma}{\gamma (\lambda-1)}\Big[1+  \left(\frac{\pi\ell_\eta}{W}\right)^2\Big]; \qquad     \ell_\kappa \xrightarrow{}\infty
\label{eq:alphac-weak}
\end{equation}
In the case of weak anchoring, this simple argument provides a good estimate for the transition from the quiescent state to the coherent laminar flow (Fig.~\ref{fig:PhaseWeakAnchor}).
On the other hand, this one dimensional model fails to account for the onset of the vortex lattice with strong anchoring.
\begin{figure}[h]
    \centering
    \begin{subfigure}[b]{0.9\linewidth}
         \centering
         \includegraphics[width=\textwidth]{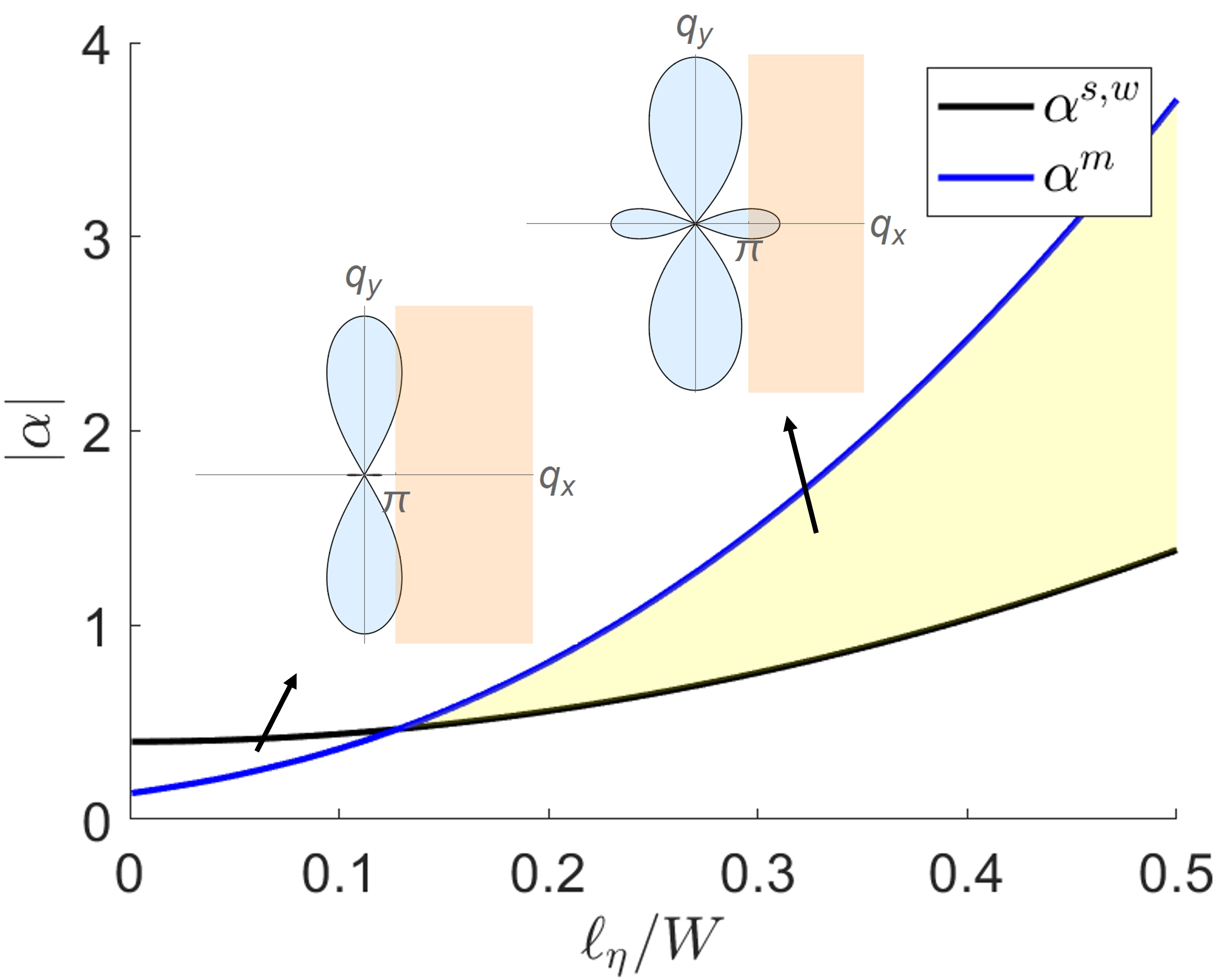}
         \caption{Weak Anchoring ($\ell_\kappa \xrightarrow{} \infty$)}
         \label{}
     \end{subfigure}
     \hfill
     \begin{subfigure}[b]{0.9\linewidth}
         \centering
         \includegraphics[width=\textwidth]{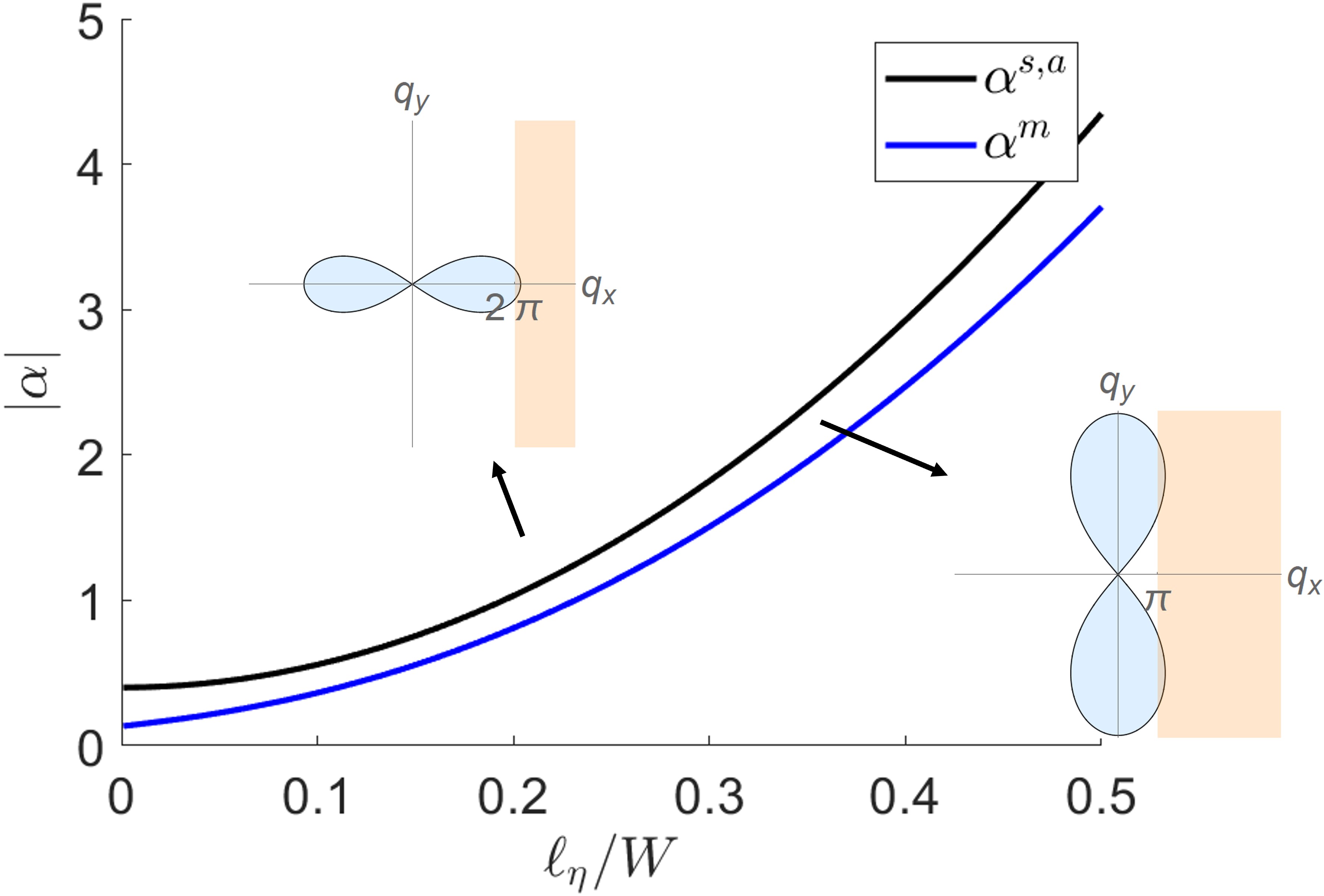}
         \caption{Strong Anchoring ($\ell_\kappa = 0$)}
         \label{}
     \end{subfigure}
     \caption{Linear instability for splay and the vortex-lattice state in the limits of weak and strong anchoring. The yellow shaded region marks where $\alpha_c^{s,w} < \alpha_c^{m}$. For the strong anchoring limit, $\alpha_c^{m}$ is always smaller than $\alpha_c^{s,a}$ and no coherent flow is possible.}
     \label{fig:LinearStability}
\end{figure}

The transition to the vortex lattice can be described as arising from the instability of a mixed bend-splay mode where both $q_x, q_y \ne 0$. To capture this instability, we fix the transverse wave number as $q_x=q_x^*$ and then determine the critical activity above which the eigenvalue $\nu(q_x=q_x^*, q_y)$ becomes positive. Fig.~\ref{fig:PhaseDiagramAnchoringStrength} indicates that the transition to the vortex lattice state depends only weakly on the strength of anchoring. For this reason we simply take $q_x^*=\pi/W$, as suggested by the fact that vortices typically have the size of the channel width at onset. The resulting critical activity for the transition to the vortex lattice state (which we refer to as a mixed instability and denote by $\alpha_c^{m}$) is compared in Fig.~\ref{fig:LinearStability} to the splay lines given by Eqs.~\eqref{eq:alphac-strong} and~\eqref{eq:alphac-weak} for the case of  strong and weak anchoring, respectively.  For strong anchoring, $\alpha_c^{m}$ is always \emph{below}  the splay line and the systems transitions directly from the quiescent state to the vortex lattice. For weak anchoring, $\alpha_c^{m}$ falls \emph{above} the splay line at large $\ell_\eta$, allowing for a region of laminar flow.

\section{Self Propulsion}
\label{sec:polar} 
Throughout the discussion we have described the bacterial fluid using a polar order parameter, $\mathbf{p}$ but the model considered has nematic symmetry. In this case the flow  arises from spontaneous symmetry breaking of the quiescent state  and is entirely determined by deformations of the polarization field. The direction of flow is equally likely to be  up or down the channel, with associated  `splay-in' and `splay-out' configurations of  polarization. In the simulations with finite anchoring this symmetry is broken externally by the boundary conditions on the polarization. The presence of a frictional substrate can, however, allow an additional propulsive force linear in $\mathbf{p}$ to be present in force balance \cite{maitra2020swimmer,brotto2013hydrodynamics,kumar2014flocking} which takes the form
\begin{equation}
    \Gamma(\mathbf{v} - v_0 \mathbf{p})= - \boldsymbol{\nabla}\Pi+
    \eta \nabla^2 \mathbf{v} + \boldsymbol{\nabla}\cdot\big( \boldsymbol{\sigma}^\text{a} +  \boldsymbol{\sigma}^\text{lc}\big)\;.
    \label{eq:v}
\end{equation}
This polar active force explicitly breaks the up-down symmetry of the flow. The homogeneous state in bulk is now a uniformly flowing state with $\mathbf{v}=v_0\mathbf{p}$. 
As a result, any finite self-propulsion $v_0$ then breaks the flow symmetry and selects the direction of the flowing state. A small self-propulsion provides therefore the minimal forcing required for creating sustained unidirectional channel flows. For small values of $v_0$, corresponding to $v_0 \ll |\alpha_0|/\Gamma\ell_\eta$, the dipolar active stress dominates over the propulsive force and the structure of the steady state flows is qualitatively unaffected. 
Above the critical activity for spontaneous flow, we recover
coherent laminar flows arising from polarization splay for weak anchoring and  flow vortex lattices for strong anchoring, as in the absence of polar self propulsion. The flow lattice, however, acquires a steady drift at speed proportional to $v_0$.
\section{Discussion} \label{sec:discussion}
Using a hydrodynamic model of extensile polar active matter,  we have examined the role of confinement and boundary alignment in controlling the spatial and temporal organizations  of active flows  in a channel.
We show that surface anchoring controls the flow structures by selectively frustrating bend or splay distortions in the polarization and that 
flows with finite throughput can only be obtained with weak surface anchoring. Strong surface anchoring leads to the formation of lattices of flow vortices, with the number of vortices determined by the aspect ratio of the channel, consistent with previous results \cite{chandragiri2019active,shendruk2017dancing}.

\begin{figure}[h]
    \centering
    \includegraphics[width=0.45\textwidth]{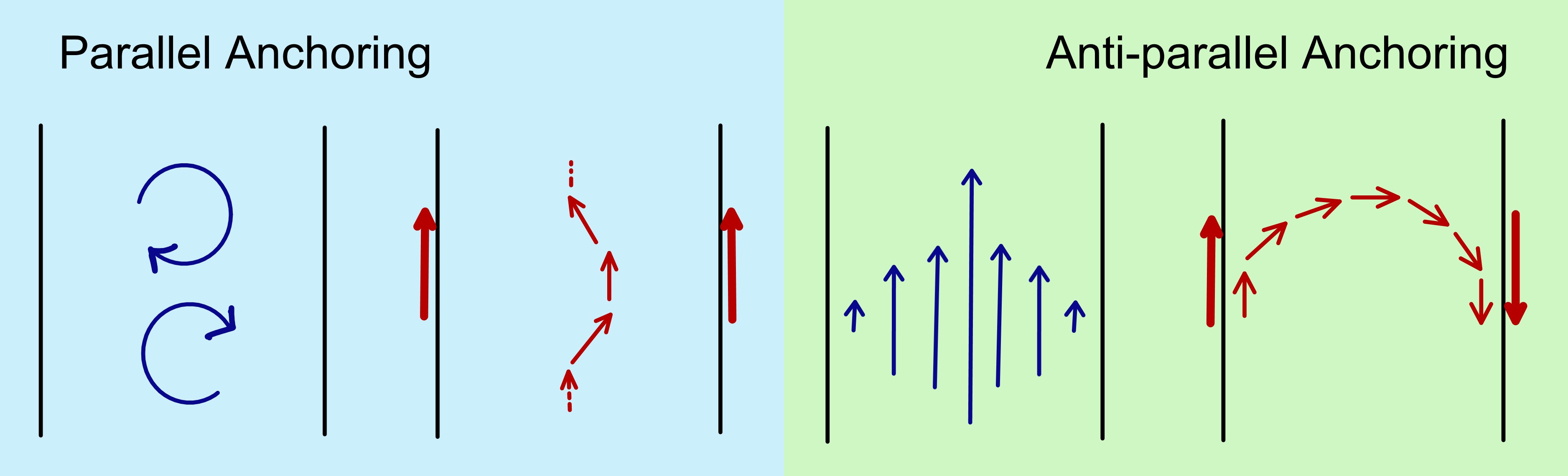}
    \caption{The symmetry of the boundary conditions for the polarization controls the steady flow states.  Left frame: strong polar anchoring where the polarization is anchored in the same direction on both channel walls suppresses splay deformations and yields bend deformation along the wall with associated flow vortices. Right frame: strong apolar anchoring where the polarization is anchored in opposite directions on the two walls  explicitly breaks the nematic symmetry  even in the absence of polar self propulsion ($v_0=0$) and leads to coherent flow. Both flow states can occur if the boundary conditions are nematic, i.e., agnostic to the direction of the polarization vector.}
    \label{fig:FlowSketchesAntiPolar}
\end{figure}

Our hydrodynamic model is inherently polar as the anchoring boundary conditions (Eq.~\ref{eq:polarAnchoring}) imposed at the channel walls break nematic symmetry even in the absence of any polar self propulsion (Sec.~\ref{sec:polar}).
As is evident, the symmetry of the boundary conditions plays a profound role on the selection of flow states. Strong polar anchoring, where the polarization is forced to point in the same direction on both sides of the channel, prevent splay deformation and facilitate bend along the channel walls, resulting in a state of flow vortices (Fig.~\ref{fig:FlowSketchesAntiPolar}, left panel). Strong antipolar anchoring, where the polarization is forced to point in opposite directions on the two sides of the channel, allows bend deformation across the channel, resulting in finite-throughput laminar flow (Fig.~\ref{fig:FlowSketchesAntiPolar}, right panel). We have also examined the case of nematic anchoring where the polarization is forced to orient with the channel wall, but with no preferred direction. This was enforced by requiring
\begin{equation}
    \left[ (\mathbf{p}\cdot\hat{\mathbf{n}})\hat{\mathbf{n}}+ \ell_\kappa (\hat{\mathbf{n}}\cdot{\bm \nabla}\mathbf{p})\right] _{x=0, W} =0\;.
\end{equation}
In this case both flow states depicted in Fig.~\ref{fig:FlowSketchesAntiPolar} can occur with equal probability, in other words they coexist in the phase diagram.

Our work quantifies the role 
surface anchoring in controlling the spatio-temporal structure of confined active flows.  
This understanding can be  useful for the design of active microfluidic devices where   channel dimensions  and boundary preparation can be independently tuned to control  viscous screening and anchoring, respectively.

Recent experiments on microtubule suspensions in 3D and associated numerical studies have demonstrated that coherent flow is only possible for finely tuned geometries~ \cite{wu2017transition,chandragiri2020flow, chandrakar2020confinement, varghese2020confinement}. Our work suggests that it  would be interesting to additionally explore the role of anchoring in these 3D systems where antagonistic boundary conditions on the three walls could result in as of yet unexplored states.  

Finally, it would also be interesting to explore the role of surface anchoring on temporal as well as spatial organization of active flows. Recent experiments in dense bacterial suspensions \cite{liu2021viscoelastic} have revealed that viscoelasticity of the suspending medium can drive a circular droplet to self-organize in time-periodic states of vortical flow, consisting of a system-spanning  vortex that switches its chirality at a rate controlled by the solvent relaxation time. 
The vortex state of a circular drops corresponds to unidirectional laminar flow in a channel. As we have seen, the direction of the flow is directly determined by the splay-in or splay-out configuration of the polarization field, which in turn can be controlled with suitable anchoring. This suggest that anchoring may play an important role in controlling temporal as well as spatial organization and that it may be possible to control oscillations between flows of opposite chirality by tuning the boundary conditions. These questions are left for future studies.

\section*{Acknowledgements} \label{sec:acknowledgements}
\noindent 
This work was directly supported by NSF grant DMR-2041459.  Use was made of computational facilities purchased with funds from the National Science Foundation (CNS-1725797) and administered by the Center for Scientific Computing (CSC). The CSC is supported by the California NanoSystems Institute and the Materials Research Science and Engineering Center (MRSEC; NSF DMR 1720256) at UC Santa Barbara. SS acknowledges support from the Harvard Society of Fellows.

\bibliographystyle{unsrt}
\bibliography{ref.bib}

\begin{thebibliography}{10}

\bibitem{wensink2012meso}
Henricus~H Wensink, J{\"o}rn Dunkel, Sebastian Heidenreich, Knut Drescher,
  Raymond~E Goldstein, Hartmut L{\"o}wen, and Julia~M Yeomans.
\newblock Meso-scale turbulence in living fluids.
\newblock {\em Proceedings of the national academy of sciences},
  109(36):14308--14313, 2012.

\bibitem{zhou2014living}
Shuang Zhou, Andrey Sokolov, Oleg~D Lavrentovich, and Igor~S Aranson.
\newblock Living liquid crystals.
\newblock {\em Proceedings of the National Academy of Sciences},
  111(4):1265--1270, 2014.

\bibitem{sanchez2012spontaneous}
Tim Sanchez, Daniel~TN Chen, Stephen~J DeCamp, Michael Heymann, and Zvonimir
  Dogic.
\newblock Spontaneous motion in hierarchically assembled active matter.
\newblock {\em Nature}, 491(7424):431--434, 2012.

\bibitem{schaller2010polar}
Volker Schaller, Christoph Weber, Christine Semmrich, Erwin Frey, and Andreas~R
  Bausch.
\newblock Polar patterns of driven filaments.
\newblock {\em Nature}, 467(7311):73--77, 2010.

\bibitem{palacci2013living}
Jeremie Palacci, Stefano Sacanna, Asher~Preska Steinberg, David~J Pine, and
  Paul~M Chaikin.
\newblock Living crystals of light-activated colloidal surfers.
\newblock {\em Science}, 339(6122):936--940, 2013.

\bibitem{bricard2013emergence}
Antoine Bricard, Jean-Baptiste Caussin, Nicolas Desreumaux, Olivier Dauchot,
  and Denis Bartolo.
\newblock Emergence of macroscopic directed motion in populations of motile
  colloids.
\newblock {\em Nature}, 503(7474):95--98, 2013.

\bibitem{marchetti2013hydrodynamics}
M~Cristina Marchetti, Jean-Fran{\c{c}}ois Joanny, Sriram Ramaswamy,
  Tanniemola~B Liverpool, Jacques Prost, Madan Rao, and R~Aditi Simha.
\newblock Hydrodynamics of soft active matter.
\newblock {\em Reviews of Modern Physics}, 85(3):1143, 2013.

\bibitem{simha2002hydrodynamic}
R~Aditi Simha and Sriram Ramaswamy.
\newblock Hydrodynamic fluctuations and instabilities in ordered suspensions of
  self-propelled particles.
\newblock {\em Physical review letters}, 89(5):058101, 2002.

\bibitem{alert2021active}
Ricard Alert, Jaume Casademunt, and Jean-Fran{\c{c}}ois Joanny.
\newblock Active turbulence.
\newblock {\em Annual Review of Condensed Matter Physics}, 13, 2021.

\bibitem{duclos2014perfect}
Guillaume Duclos, Simon Garcia, HG~Yevick, and P~Silberzan.
\newblock Perfect nematic order in confined monolayers of spindle-shaped cells.
\newblock {\em Soft matter}, 10(14):2346--2353, 2014.

\bibitem{doostmohammadi2016stabilization}
Amin Doostmohammadi, Michael~F Adamer, Sumesh~P Thampi, and Julia~M Yeomans.
\newblock Stabilization of active matter by flow-vortex lattices and defect
  ordering.
\newblock {\em Nature communications}, 7(1):1--9, 2016.

\bibitem{duclos2018spontaneous}
G~Duclos, C~Blanch-Mercader, V~Yashunsky, G~Salbreux, J-F Joanny, J~Prost, and
  Pascal Silberzan.
\newblock Spontaneous shear flow in confined cellular nematics.
\newblock {\em Nature physics}, 14(7):728--732, 2018.

\bibitem{thijssen2021submersed}
Kristian Thijssen, Dimitrius~A Khaladj, S~Ali Aghvami, Mohamed~Amine Gharbi,
  Seth Fraden, Julia~M Yeomans, Linda~S Hirst, and Tyler~N Shendruk.
\newblock Submersed micropatterned structures control active nematic flow,
  topology, and concentration.
\newblock {\em Proceedings of the National Academy of Sciences}, 118(38), 2021.

\bibitem{voituriez2005spontaneous}
R~Voituriez, Jean-Fran{\c{c}}ois Joanny, and Jacques Prost.
\newblock Spontaneous flow transition in active polar gels.
\newblock {\em EPL (Europhysics Letters)}, 70(3):404, 2005.

\bibitem{wioland2013confinement}
Hugo Wioland, Francis~G Woodhouse, J{\"o}rn Dunkel, John~O Kessler, and
  Raymond~E Goldstein.
\newblock Confinement stabilizes a bacterial suspension into a spiral vortex.
\newblock {\em Physical review letters}, 110(26):268102, 2013.

\bibitem{lushi2014fluid}
Enkeleida Lushi, Hugo Wioland, and Raymond~E Goldstein.
\newblock Fluid flows created by swimming bacteria drive self-organization in
  confined suspensions.
\newblock {\em Proceedings of the National Academy of Sciences},
  111(27):9733--9738, 2014.

\bibitem{wu2017transition}
Kun-Ta Wu, Jean~Bernard Hishamunda, Daniel~TN Chen, Stephen~J DeCamp, Ya-Wen
  Chang, Alberto Fern{\'a}ndez-Nieves, Seth Fraden, and Zvonimir Dogic.
\newblock Transition from turbulent to coherent flows in confined
  three-dimensional active fluids.
\newblock {\em Science}, 355(6331), 2017.

\bibitem{chen2018dynamics}
Sheng Chen, Peng Gao, and Tong Gao.
\newblock Dynamics and structure of an apolar active suspension in an annulus.
\newblock {\em Journal of Fluid Mechanics}, 835:393--405, 2018.

\bibitem{opathalage2019self}
Achini Opathalage, Michael~M Norton, Michael~PN Juniper, Blake Langeslay, S~Ali
  Aghvami, Seth Fraden, and Zvonimir Dogic.
\newblock Self-organized dynamics and the transition to turbulence of confined
  active nematics.
\newblock {\em Proceedings of the National Academy of Sciences},
  116(11):4788--4797, 2019.

\bibitem{you2021confinement}
Zhihong You, Daniel~JG Pearce, and Luca Giomi.
\newblock Confinement-induced self-organization in growing bacterial colonies.
\newblock {\em Science Advances}, 7(4):eabc8685, 2021.

\bibitem{wioland2016directed}
Hugo Wioland, Enkeleida Lushi, and Raymond~E Goldstein.
\newblock Directed collective motion of bacteria under channel confinement.
\newblock {\em New Journal of Physics}, 18(7):075002, 2016.

\bibitem{tjhung2011nonequilibrium}
Elsen Tjhung, Michael~E Cates, and Davide Marenduzzo.
\newblock Nonequilibrium steady states in polar active fluids.
\newblock {\em Soft Matter}, 7(16):7453--7464, 2011.

\bibitem{giomi2008complex}
Luca Giomi, M~Cristina Marchetti, and Tanniemola~B Liverpool.
\newblock Complex spontaneous flows and concentration banding in active polar
  films.
\newblock {\em Physical review letters}, 101(19):198101, 2008.

\bibitem{yang2016role}
Xiaogang Yang and Qi~Wang.
\newblock Role of the active viscosity and self-propelling speed in channel
  flows of active polar liquid crystals.
\newblock {\em Soft Matter}, 12(4):1262--1278, 2016.

\bibitem{chandragiri2019active}
Santhan Chandragiri, Amin Doostmohammadi, Julia~M Yeomans, and Sumesh~P Thampi.
\newblock Active transport in a channel: stabilisation by flow or
  thermodynamics.
\newblock {\em Soft matter}, 15(7):1597--1604, 2019.

\bibitem{shendruk2017dancing}
Tyler~N Shendruk, Amin Doostmohammadi, Kristian Thijssen, and Julia~M Yeomans.
\newblock Dancing disclinations in confined active nematics.
\newblock {\em Soft Matter}, 13(21):3853--3862, 2017.

\bibitem{samui2021flow}
Abhik Samui, Julia~M Yeomans, and Sumesh~P Thampi.
\newblock Flow transitions and length scales of a channel-confined active
  nematic.
\newblock {\em Soft Matter}, 17(47):10640--10648, 2021.

\bibitem{wagner2022exact}
Caleb~G Wagner, Michael~M Norton, Jae~Sung Park, and Piyush Grover.
\newblock Exact coherent structures and phase space geometry of preturbulent 2d
  active nematic channel flow.
\newblock {\em Physical Review Letters}, 128(2):028003, 2022.

\bibitem{hardouin2019reconfigurable}
J{\'e}r{\^o}me Hardo{\"u}in, Rian Hughes, Amin Doostmohammadi, Justine Laurent,
  Teresa Lopez-Leon, Julia~M Yeomans, Jordi Ign{\'e}s-Mullol, and Francesc
  Sagu{\'e}s.
\newblock Reconfigurable flows and defect landscape of confined active
  nematics.
\newblock {\em Communications Physics}, 2(1):1--9, 2019.

\bibitem{norton2018insensitivity}
Michael~M Norton, Arvind Baskaran, Achini Opathalage, Blake Langeslay, Seth
  Fraden, Aparna Baskaran, and Michael~F Hagan.
\newblock Insensitivity of active nematic liquid crystal dynamics to
  topological constraints.
\newblock {\em Physical Review E}, 97(1):012702, 2018.

\bibitem{rorai2021active}
Cecilia Rorai, Federico Toschi, and Ignacio Pagonabarraga.
\newblock Active nematic flows confined in a two-dimensional channel with
  hybrid alignment at the walls: A unified picture.
\newblock {\em Physical Review Fluids}, 6(11):113302, 2021.

\bibitem{poujade2007collective}
Mathieu Poujade, Erwan Grasland-Mongrain, A~Hertzog, J~Jouanneau, Philippe
  Chavrier, Beno{\^\i}t Ladoux, Axel Buguin, and Pascal Silberzan.
\newblock Collective migration of an epithelial monolayer in response to a
  model wound.
\newblock {\em Proceedings of the National Academy of Sciences},
  104(41):15988--15993, 2007.

\bibitem{conrad2018confined}
Jacinta~C Conrad and Ryan Poling-Skutvik.
\newblock Confined flow: consequences and implications for bacteria and
  biofilms.
\newblock {\em Annual review of chemical and biomolecular engineering},
  9:175--200, 2018.

\bibitem{beebe2002physics}
David~J Beebe, Glennys~A Mensing, and Glenn~M Walker.
\newblock Physics and applications of microfluidics in biology.
\newblock {\em Annual review of biomedical engineering}, 4(1):261--286, 2002.

\bibitem{clark2015modes}
Andrew~G Clark and Danijela~Matic Vignjevic.
\newblock Modes of cancer cell invasion and the role of the microenvironment.
\newblock {\em Current opinion in cell biology}, 36:13--22, 2015.

\bibitem{needleman2017active}
Daniel Needleman and Zvonimir Dogic.
\newblock Active matter at the interface between materials science and cell
  biology.
\newblock {\em Nature reviews materials}, 2(9):1--14, 2017.

\bibitem{degennes}
P.G. DeGennes and J.~Prost.
\newblock {\em Physics of Liquid Crystals}.
\newblock Clarendon, Oxford, 1994.

\bibitem{giomi2015geometry}
Luca Giomi.
\newblock Geometry and topology of turbulence in active nematics.
\newblock {\em Physical Review X}, 5(3):031003, 2015.

\bibitem{hemingway2016correlation}
Ewan~J Hemingway, Prashant Mishra, M~Cristina Marchetti, and Suzanne~M
  Fielding.
\newblock Correlation lengths in hydrodynamic models of active nematics.
\newblock {\em Soft Matter}, 12(38):7943--7952, 2016.

\bibitem{maitra2020swimmer}
Ananyo Maitra, Pragya Srivastava, M~Cristina Marchetti, Sriram Ramaswamy, and
  Martin Lenz.
\newblock Swimmer suspensions on substrates: anomalous stability and long-range
  order.
\newblock {\em Physical review letters}, 124(2):028002, 2020.

\bibitem{alnaes2015fenics}
Martin Aln{\ae}s, Jan Blechta, Johan Hake, August Johansson, Benjamin Kehlet,
  Anders Logg, Chris Richardson, Johannes Ring, Marie~E Rognes, and Garth~N
  Wells.
\newblock The fenics project version 1.5.
\newblock {\em Archive of Numerical Software}, 3(100), 2015.

\bibitem{logg2012automated}
Anders Logg, Kent-Andre Mardal, and Garth Wells.
\newblock {\em Automated solution of differential equations by the finite
  element method: The FEniCS book}, volume~84.
\newblock Springer Science \& Business Media, 2012.

\bibitem{green2017geometry}
Richard Green, John Toner, and Vincenzo Vitelli.
\newblock Geometry of thresholdless active flow in nematic microfluidics.
\newblock {\em Physical Review Fluids}, 2(10):104201, 2017.

\bibitem{brotto2013hydrodynamics}
Tommaso Brotto, Jean-Baptiste Caussin, Eric Lauga, and Denis Bartolo.
\newblock Hydrodynamics of confined active fluids.
\newblock {\em Physical review letters}, 110(3):038101, 2013.

\bibitem{kumar2014flocking}
Nitin Kumar, Harsh Soni, Sriram Ramaswamy, and AK~Sood.
\newblock Flocking at a distance in active granular matter.
\newblock {\em Nature communications}, 5(1):1--9, 2014.

\bibitem{chandragiri2020flow}
Santhan Chandragiri, Amin Doostmohammadi, Julia~M Yeomans, and Sumesh~P Thampi.
\newblock Flow states and transitions of an active nematic in a
  three-dimensional channel.
\newblock {\em Physical Review Letters}, 125(14):148002, 2020.

\bibitem{chandrakar2020confinement}
Pooja Chandrakar, Minu Varghese, S~Ali Aghvami, Aparna Baskaran, Zvonimir
  Dogic, and Guillaume Duclos.
\newblock Confinement controls the bend instability of three-dimensional active
  liquid crystals.
\newblock {\em Physical Review Letters}, 125(25):257801, 2020.

\bibitem{varghese2020confinement}
Minu Varghese, Arvind Baskaran, Michael~F Hagan, and Aparna Baskaran.
\newblock Confinement-induced self-pumping in 3d active fluids.
\newblock {\em Physical Review Letters}, 125(26):268003, 2020.

\bibitem{liu2021viscoelastic}
Song Liu, Suraj Shankar, M~Cristina Marchetti, and Yilin Wu.
\newblock Viscoelastic control of spatiotemporal order in bacterial active
  matter.
\newblock {\em Nature}, 590(7844):80--84, 2021.

\end{thebibliography}

\newpage
\appendix
\section{Elastic Energy} \label{sec:appendixElasticEnergy}
In this Appendix we use simple energetic arguments to estimate the scaling of the number of vortices obtained in the vortex state with the channel aspect ratio.
Both the shear banded and the vortex lattice states correspond to configurations where $\abs{\mathbf{p}} \approx 1$ and the main deformation mode is a bend distortion in the polarization field,  described by the angle $\theta$, with $\cos{\theta} = \mathbf{p}\cdot\mathbf{\hat{x}}$.
\begin{figure}[h]
\centering
    \begin{subfigure}[b]{0.9\linewidth}
         \centering
         \includegraphics[width=\textwidth]{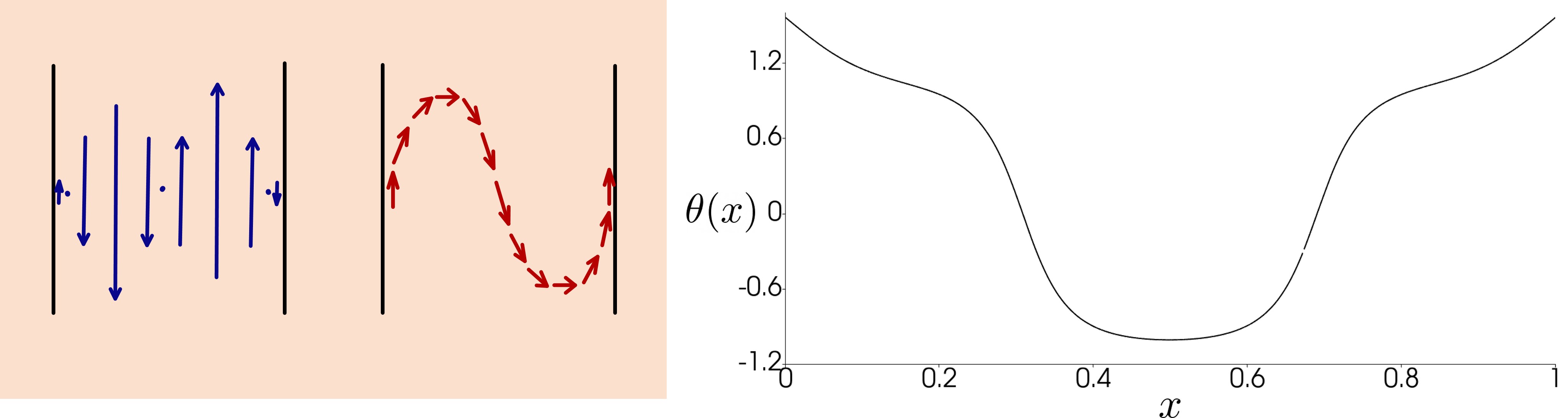}
         \caption{Shear Banded Flow (angular profile is shown for $\alpha =-3.80$).}
         \label{}
     \end{subfigure}
     \hfill
     \begin{subfigure}[b]{0.9\linewidth}
         \centering
         \includegraphics[width=\textwidth]{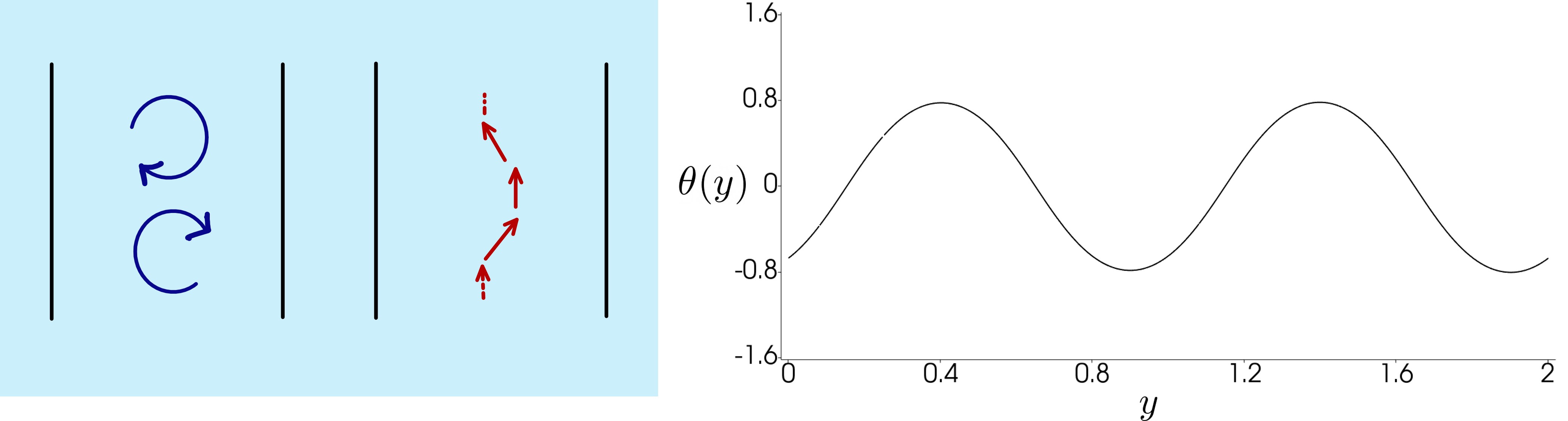}
         \caption{Flow Vortex Lattice (angular profile is shown for $\alpha =-2.00$).}
         \label{}
     \end{subfigure}
     \caption{Schematic (L=W) for the shear banded and the vortex lattice flows. The line plot shows the angle $\theta$ along the polarization $\mathbf{p}$ for the two flow states from the numerical simulations (L=2W) in the case of strong anchoring ($\ell_\kappa/W = 0.01, \ell_\eta/W =0.30$). }
    \label{fig:SketchLinePlots}
\end{figure}

Upon transition from the quiescent to the vortex state, the flow  organizes into a single pair of counter-rotating vortices  spanning the channel length $L$ and of size of order $L/2$ (provided $L\geq W$),
as illustrated in Fig. \ref{fig:Profiles_Strong} for $\alpha =-2$. This is observed in simulations for a large range of $\ell_\eta$ and $\ell_\kappa.$
Upon increasing activity, the   number of vortex pairs increases, until eventually the system transitions to shear banded flow. The maximum number $n$ of vortex pairs that can be accommodated in the channel depends of the channel's aspect ratio. To estimate this number, we examine the elastic energy of channel-spanning deformations, given by
\begin{equation}
    E \approx K\int_\Omega \mathrm{d}\mathbf{x} \, (\nabla \theta)^2\;. 
    \label{eq:Etheta}
\end{equation}
In the shear banded  state, the bend deformation is primarily transverse to the channel direction, corresponding to an angle profile of the form
\begin{equation}
    \theta_s \sim A_s\sin(2\pi x/W)\;.
    \label{eq:thetas}
\end{equation}
In the vortex lattice  state, away from the walls,  bend deformations are primarily along the length of the channel, corresponding to an angle profile of the form 
\begin{equation}
    \theta_v \sim A_v\sin(2\pi n y/L)\;,
    \label{eq:thetav}
\end{equation}
where $n$ is the number of counter-rotating vortex pairs. 
Note that the amplitudes $A_s$ and $A_v$ of the two deformations depend on activity and on the strength of anchoring, and are generally different.

The corresponding deformation energies $E_s$ and $E_v$ for the shear banded and flow states are then immediately obtained by substituting \eqref{eq:thetas} and \eqref{eq:thetav} into \eqref{eq:Etheta}, with the result
\begin{align}
    E_s &= K A_s^2L/W \;,\\
    E_v &= K n^2 A_v^2 W/L\;.
\end{align}
By setting $E_s\sim E_v$ we can estimate the number of vortex pairs in the channel as
\begin{equation}
    n \sim (L/W) (A_s/A_v)\;.
    \label{eq:nv}
\end{equation}

The scaling of the maximum number of  vortex pairs with the channel aspect ratio at the transition between a lattice of flow vortices and the shear banded flow is confirmed by numerical simulations in longer channels.

For weak anchoring, the direction of polarization $\theta$ is essentially free at the channel walls. The shear banded flow always cost more energy than the vortex state, resulting in $A_s<A_v$ (illustrated in the schematic in Fig. \ref{fig:FlowSketches} in the two anchoring limits). 
A vortex lattice can appear, but it is transient and the steady state is stable shear banded flow.

\end{document}